\documentclass[fleqn,usenatbib]{mnras}

\usepackage{newtxtext,newtxmath}
\usepackage[T1]{fontenc}

\DeclareRobustCommand{\VAN}[3]{#2}
\let\VANthebibliography\thebibliography
\def\thebibliography{\DeclareRobustCommand{\VAN}[3]{##3}\VANthebibliography}

\usepackage{graphicx}	% Including figure files
\usepackage{amsmath}	% Advanced maths commands
\usepackage{tabularx}
\usepackage[cp866]{inputenc}
\usepackage{supertabular}
\newcommand*\rot{\rotatebox{90}}

\title{
Pushchino multibeam pulsar search VI. Method of pulsar timing using bad timed data
}
\author[Andrianov et al.]{S.~A. Andrianov, $^{1}$\thanks{E-mail: andrianovs@prao.ru}
V. A. Potapov,$^{1}$
S. A. Tyul'bashev,$^{1}$
S.V. Logvinenko,$^{1}$
V. V. Oreshko,$^{1}$
\\
$^{1}$ Lebedev Physical Institute, Astro Space Center, Pushchino Radio Astronomy Observatory,  
Russian Academy of Sciences (PRAO ASC LPI), Pushchino, Moscow reg., 142290 Russia. 
}%
\date{2024}

\pubyear{2024}

\begin{document}
\label{firstpage}
\pagerange{\pageref{firstpage}--\pageref{lastpage}}
\maketitle
\begin{abstract}
A method for pulsar timing  based on monitoring data from the 3-th diagramm of the Large Phase Array (LPA LPI) radio telescope is proposed. In our observations, recorders with quartz clock generators were used as local clocks. Such recorders initially had an accuracy and hardware reference to the UTC time scale insufficient for pulsar timing. We have developed a method for referencing such clocks to the UTC based on observations of known pulsars used as intermediate reference clocks. This allowed us to improve dramatically the accuracy of determining the Time of Arrivals (TOA) of pulsars' pulses.

We applied this method to the results of our observations of 24 second period pulsars over a time interval of 10 years. It was shown that the accuracy of the pulsar period, its first derivative ($P$ and $\dot P$) and their coordinates in right ascension and declination ($\alpha, \delta$) allow us to predict the pulsar phase within $\pm 0.5 P$ during several years. The accuracy of determining the coordinates by right ascension and declination was typically better than $10^{\prime \prime}$ with an angular resolution of the radio telescope of about $30^\prime$. That makes it possible to use these parameters for timing using radio telescopes with narrow beam patterns. The accuracy of the calculated period was typically better than $10^{-8}$~s.
\end{abstract}


\section{Introduction}

Pulsars were discovered by their periodic emission in 1967 (\citeauthor{Hewish1968}, \citeyear{Hewish1968}). The periods of own rotation of pulsars ($P$) are stable and change slowly over time. According to the ATNF CSIRO pulsar catalog (\citeauthor{Manchester2005}, \citeyear{Manchester2005})\footnote{https://www.atnf.csiro.au/research/pulsar/psrcat /}, the increase in the period of a typical second (slow, with a self-rotation period of $> 100$~ms) pulsar over a year of observations ranges from $10^{-6}$ to $10^{-9}$s, which corresponds to the range of values of its derivative ($\dot P \sim 10^{-13} - 10^{-16}$~ s/s). This increase in the period is mainly due to the loss of rotational energy of pulsars to electromagnetic radiation (magnetodipole losses). At the same time, the period value is often determined in catalogs better than $10^{-12}$~s. The time to determine the Time of arrival (TOA) of a single pulsar observation at the barycenter of the Solar System varies widely from $10^{-2}$ to $10^{-7}$ ~s, depending on the type of pulsar, radio frequency and quality of observations (\citeauthor{Lorimer2008}, \citeyear{Lorimer2008}).

There is a class of astrophysical problems that can be solved only with the help of high-precision pulsar timing, when high accuracy of determining the TOA of a pulsar, its period and its derivatives may be required. These include, for example, the classical problem of determining the Keplerian and post-Keplerian (relativistic) parameters of binary systems based on observations of their pulsars (\citeauthor{Damour1986}, \citeyear{Damour1986}), the tasks of testing general relativity and alternative theories of gravity (see for example, reviews of (\citeauthor{Damour1992}, \citeyear{Damour1992}; \citeauthor{Kramer2021}, \citeyear{Kramer2021})), the task of searching for gravitational wave radiation and/or determining its upper limit (\citeauthor{Kopeikin2004}, \citeyear{Kopeikin2004}; \citeauthor{Potapov2003}, \citeyear{Potapov2003}; \citeauthor{Arzoumanian2021}, \citeyear{Arzoumanian2021}). Adjacent to these tasks is the task of determining the fundamental scale of astronomical Pulsar Time (PT) (\citeauthor{Zharov2019}, \citeyear{Zharov2019}). The high accuracy of the TOA determination can also be useful in the tasks of studying the interstellar medium, in particular, the secular changes in electron density on the line of sight of the pulsar observer (\citeauthor{Ilyasov2005}, \citeyear{Ilyasov2005}).

One of the most famous results of high-precision timing is the study of the B1913+16 millisecond binary pulsar, which made it possible to conduct a number of general relativity tests, in particular, to determine an additional increase in its orbital period associated with energy losses due to gravitational wave radiation. At the same time, the error in determining the pulsar period was $10^{-14}$~s.

Despite the fact that both the accuracy of determining pulsar parameters and the accuracy of timing in general (defined here as r.m.s. of residual deviations (residual) of TOA, i.e. TOA residuals of pulsars, see Chapter 4 below) during observations of slow pulsars is significantly, typically by 2 orders of magnitude, inferior to that of millisecond pulsars, a number of tasks can be listed that are successfully solved with the help of their observations.

For example, the above-mentioned accuracy of determining the period is excessive for observations of slow pulsars, which aim to obtain an average profile and determine its main energy characteristics when the observation session lasts from minutes to tens of minutes ($\sim 10^{2} - 10^{3}$ pulses), and for signal accumulation with satisfactory accuracy, it is sufficient to know $P$ up to the fourth or fifth decimal place. Such tasks also include: the study of statistics on changes in the pulse flux density, in particular, the study of anomalous and giant pulses (GP) (\citeauthor{Kazantsev2018}, \citeyear{Kazantsev2018}), the study of the interstellar medium: its turbulence and short-term variations in electron density (\citeauthor{Shishov2003}, \citeyear{Shishov2003}; \citeauthor{Losovsky2024}, \citeyear{Losovsky2024}).

A pulsar period failure (glitch) looks like a short-term change in the period followed by its recovery (glitch of the first type) or a constant shift (glitch of the second type) while maintaining the value of the derivative of the period. The glitch values for most pulsars are in the range of $\dot P/P\sim 10^{-6} - 10^{-11}$  (\citeauthor{Espinoza2011}, \citeyear{Espinoza2011}), which for glitches of the second type will result in a change in the observed TOA by values from tens of microseconds to seconds per year.

The timing of slow radio pulsars can be used in the task of spacecraft navigation in deep space. To determine the current coordinates of the spacecraft, it is possible to measure the TOA and/or changes in the periods of the pulsars with an on-board antenna, which makes it possible to determine their position relative to the initial coordinate of the spacecraft and the current velocity. The higher the accuracy of determining the spacecraft's location and velocity, the more accurately we can estimate the TOA and the pulsar period. However, the limitations on the flux density associated with the small effective area of the on-board antenna, the steeply (usually as $\sim 1 /f^2$) increasing pulsar spectra to low radio frequencies, as well as the fact that the flux density of slow pulsars is much higher than the flux density of millisecond pulsars, in practice makes it possible to use only the first of these (\citeauthor{Rodin2020}, \citeyear{Rodin2020}).

The review of (\citeauthor{Lorimer2004}, \citeyear{Lorimer2004}) (Fig.1.13) shows a well-known diagram of the dependence of the pulsar period on its derivative: $P/ \dot P$. This diagram clearly distinguishes between slow (second) and millisecond pulsars, binary pulsars, X-ray and gamma pulsars, and radio quite pulsars. Relatively recently discovered rotating radio transients (\citeauthor{McLaughlin2006}, \citeyear{McLaughlin2006}) and long-period transients (\citeauthor{Caleb2022}, \citeyear{Caleb2022}; \citeauthor{Hurley-Walker2022}, \citeyear{Hurley-Walker2022}) also occupy their area in this diagram (\citeauthor{Keane2008}, \citeyear{Keane2008}; \citeauthor{Cui2017}, \citeyear{Cui2017}; \citeauthor{Rea2024}, \citeyear{Rea2024}; \citeauthor{Caleb2024}, \citeyear{Caleb2024}). That is, the dependence of $P/\dot P$ allows you to immediately estimate which sample an open pulsar can fall into, without needing the extreme accuracy given above.

The accuracy of determining the parameters of pulsar timing depends on a number of main factors related to the pulsar itself: the period of the pulsar and its type, pulse duration, and related to the quality of observations: the duration and density of the data series, radio frequency, signal-to-noise ratio (S/N) in the pulse. The shorter the period, the narrower the pulse, the longer the observation period, and the higher the S/N in the profile, the better the TOA is determined and the lower the error in determining the parameters of the pulsar (\citeauthor{Lorimer2008}, \citeyear{Lorimer2008}).

Obviously, the accuracy of the TOA determination directly depends on the quality of the reference clock. Therefore, in observatories engaged in timing, the clock frequency generator of the pulsar signal recorder is synchronized with the reference frequency, which allows for long-term observation sessions, and the start time of signal recording is tied with high accuracy to the UTC time scale.

In 2014, the Pushchino Radio Astronomy Observatory (PRAO) launched daily monitoring observations, which continue to the present day. The initial purpose of monitoring is Space Weather forecasting (\citeauthor{Shishov2016}, \citeyear{Shishov2016}). The data obtained also makes it possible to search for pulsars and detect individual dispersed pulses from cosmic sources (\citeauthor{Tyulbashev2016}, \citeyear{Tyulbashev2016}; \citeauthor{Tyulbashev2018}, \citeyear{Tyulbashev2018}). According to the Pushchino Multibeam Pulsar Search project (PUMPS) discovered approximately 100 new pulsars and 100 rotating radio transients (RRAT)\footnote{https://bsa-analytics.prao.ru/}. The observation series for many pulsars entering the monitoring area has already exceeded 10 years, and the timing of pulsars and RRAT is an obvious task. However, the recorders that record observations are equipped with conventional quartz oscillators, which have unsatisfactory frequency stability for timing and an irregular time reference to the UTC reference time scale.

Next, we are considering the possibility of using the TOA we obtained in such a local scale for timing with satisfactory quality. For this purpose, slow pulsars with well-known parameters were used as intermediate clocks. The chapter \ref{Observations} describes the features of conducting observations on the 3rd diagram of the LPA radio telescope, the chapter \ref{Clocks} describes the characteristics of reference quartz oscillators, the chapter \ref{Timing} describes the method of linking the TOA obtained on the local scale to the UTC scale (using reference pulsars), and the results of its application to the TOA of 24 known pulsars observed on LPA-3 are also summarized in the chapter \ref{Discussion}. The main results of the work are summarized.

\section{Observations} \label{Observations}

A Large Phased Array (LPA) is an antenna array built on wave dipoles. Its size is approximately equal to $200 \times 400$~m. During the reconstruction of the antenna, which ended in 2013 (\citeauthor{Shishov2016}, \citeyear{Shishov2016}; \citeauthor{Tyulbashev2016}, \citeyear{Tyulbashev2016}), the distributed gain system of the phased array was replaced. The operating frequency range of the LPA is 109-113 MHz, the frequency range of the recorder is 109-111.5 MHz. Low-noise amplifiers at the output of the dipole lines have one input and four independent outputs. Therefore, it is possible to implement up to four directional patterns/radio telescopes based on a single receiving antenna (two are currently operating), controlled independently. One of these radio telescopes (LPA3) has 128 stationary (fixed) beams overlapping the meridian plane from $-9^{\circ}$ to $+55^{\circ}$ in declination ($\delta$). The intersection of the beams occurs at the level of 0.405. The size of the radiation pattern (beam) is approximately equal to $0.5^{\circ}\times 1^{\circ}$. The passage of the source through the beam at half power is approximately 3.5 minutes. Three digital recorders (48+48+32 beams) allow you to simultaneously record all 128 beams. Thus, the instantaneous viewing area is approximately 50 square meters.deg., and almost half of the celestial sphere is recorded per day.

High time-frequency resolution is not required for observations under the "Space Weather" (\citeauthor{Shishov2016}, \citeyear{Shishov2016}) project. Recording is performed in the mode of 6 frequency channels with a width of 415 kHz with a sampling of 100 ms. This mode allows us to fully observe the interplanetary scintillations of compact radio sources and predict coronal mass ejections and the arrival of corotating interaction regions to the Earth. In August 2014, an additional observation mode with a higher time-frequency resolution was implemented (\citeauthor{Tyulbashev2016}, \citeyear{Tyulbashev2016}). In this mode, the 2.5 MHz band is divided into 32 of 78 kHz wide frequency channels, with a sampling of 12.4928 ms. Data with low and high time-frequency resolution is recorded on the recorder's hard disks simultaneously.

The recorders were commissioned at different times, so the series of observations have different durations. The maximum duration of the observations is provided by the recorder, which records declinations of $+21^{\circ} < \delta < +42^{\circ}$. At the same recorders, there is a minimum of interference associated with the large industrial centers of Moscow and Tula, located about 100 km from Pushchino in the north and south.

The main purpose for the 128-ray of the LPA3 was scintillation radio sources, based on changes in the degree of scintillations of which the forecast of "Space Weather" is made. For such observations, high-precision reference to the time scale is not necessary. When designing recorders at the hardware level, the use of a quartz frequency generator as a reference was included. The recorder data is recorded in hourly blocks, and within the block, the stability of the time intervals between points is determined by the stability of the quartz oscillator. The start time of recording is tied to the UTC time scale with an accuracy of at least 2 ms.

High time-frequency resolution data is used to search for pulsars and RRAT (\citeauthor{Tyulbashev2016}, \citeyear{Tyulbashev2016}; \citeauthor{Tyulbashev2018}, \citeyear{Tyulbashev2018}). During a 3.5-minute observation session (the time the source passes through the meridian at half the power of the radiation pattern of the radio telescope), the possible time loss may be several milliseconds, which is less than the sampling interval for one point in the observations at LPA3. Therefore, the time error is not reflected in the Fourier power spectra used to search for pulsars. These errors also do not interfere with the search for RRAT pulses.

The recorder consists of modules. Each module is implemented on a programmable logic integrated circuit (FPGA) and records from eight beams. Each module has its own quartz oscillators. In laboratory conditions, it was found that the accuracy of determining the first reading when starting the generator is approximately $\pm 6$~ms. During an hour-long observation session, a clock based on a quartz oscillator for a separate module may lag or rush by 25 ms ($\pm 2$ points) (\citeauthor{Tyulbashev2016}, \citeyear{Tyulbashev2016}). In real-world observations, prior to this work, the declared (laboratory) accuracy of the module's local clock had not been verified.

\section{Features of working with quartz frequency generators on LPA} \label{Clocks}

To synchronize the system time in the observatory's local network, two GPS (Global Positioning System) receivers with built-in NTP (Network Time Protocol) servers and time services installed on local computers are used, which require higher time scale accuracy compared to conventional computers. To do this, the standard time service (W32Time, Microsoft) is replaced by a service (Network Time Protocol Daemon, Meinberg) supplied by manufacturers of GPS receivers and allowing you to synchronize the local time scale with the time of NTP servers with an error of less than 2 ms.

The process of writing data to a hard disk is controlled by an observation program and is performed in 1-hour blocks. In fact, the recording is continuous, with the peculiarity that when passing through time after a whole hour, the current file is closed and a new one is opened. This recording mode does not require the use of significant operating system resources and does not interfere with a more accurate restart of data logging. After the recording start signal is issued, the process of forming points reflecting the flux densities in the frequency channels over time begins, and recording them to disk. The formation of a time stream of spectra with the required time and frequency resolution is performed using modules installed in PCI slots of industrial computers. Each module contains eight 12-bit ADCs, digitized data streams from which are processed in parallel on the FPGA, after which the data for recording the eight beams is transferred to the computer's RAM (\citeauthor{Logvinenko2024}, \citeyear{Logvinenko2024}).

Each module has its own quartz oscillator, on the basis of which the clock frequencies necessary for digitization and signal processing are generated. It follows that, in each module, the clock frequencies are independent and have a different sampling period error. Software and hardware are used to reduce the impact of this factor on the definition of TOA.

The software includes both hourly synchronization of system time for all recorders modules using NTP servers on a local network, and software tools for determining and triggering the start of registration by an integer number of hours of the time scale. For this purpose, system timers with an accuracy of milliseconds are used. The first timer is set to trigger at a time 1000 ms behind the start of recording (the end of the hour interval). Since the timer in this case uses a quartz oscillator installed on the computer's motherboard, its error in the time interval of about an hour may be significant. This timer starts another timer that is triggered 2 ms before the end of the hour interval. The response error for a time interval of less than a second is significantly reduced. According to this timer, a restart signal is issued for all recorder modules. The restart signal for the modules is issued in a separate thread of the operating system with the Highest Priority. Thus, the start time of each hourly recorder recording is synchronized with the system time and, accordingly, with the UTC time scale.

The hardware means of increasing the accuracy of TOA detection include work to improve the quality of reference frequency generators. For the modules of the second recorder, quartz oscillators were selected according to the parameter of the true, more accurate generation frequency than indicated in the technical description. Since we had a small number of oscillator instances, the frequency spread was reduced slightly. Of course, the best option would be to purchase a batch of generators with an increased value of the accuracy parameter of the absolute frequency of generation.

As we wrote in the paragraph above, the original purpose of the recorders (modules) was to observe scintillation radio sources, and therefore, when developing the modules, it was not possible to use an external reference frequency. The requirement of an external reference frequency increases the price of the recorder, since high requirements are imposed for supplying special-purpose signals to the FPGA (ADC, PCI bus). The standard PRAO pulsar recorder connected to the LPA1 radio telescope uses external reference frequency signals from the GPS receiver and a 1 Hz (1 second) time scale synchronization signal, which allows for time reference accuracy up to $\pm 100$~ns.

Quartz reference oscillators installed on different modules have their own frequency error and its variation over time. To carry out timing work, it is necessary to investigate the timing of the loggers' output data on each of the 16 modules (16 modules $\times$8 beams = 128). In this paper, we talk about the study of 6 recorder modules serving beams with a minimum of observed industrial interference.

\section{Timing of pulsars using the local quartz clock time scale} \label{Timing}

Let us consider the possibility of timing pulsars using local reference clocks with {\it a priori} unknown course relative to the uniform UTC time scale.
The task is standard, and is reduced to several stages (see, for example, a detailed description of the algorithms for timing single pulsars in (\citeauthor{Doroshenko1990}, \citeyear{Doroshenko1990}; \citeauthor{Hobbs2006}, \citeyear{Hobbs2006}), implemented in the TIMAPR405 and TEMPO2 software packages, respectively):

\begin{itemize}
\item Obtaining an mean pulsar profile with a satisfactory signal-to-noise ratio (we consider the S/N of the mean profile $\ge 7$ to be such a criterion), for which, in observations of slow pulsars at the LPA, it is usually sufficient to synchronize the signal with the summation period during one session. The duration of the LPA observation session is determined by the time of passage through 1/2 of the antenna beam pattern ($\ge 3.5$~min.). For the strongest pulsars, even single periods can be used if the pulse profile is resolved.
\item Definition of topocentric TOA in the local (on the telescope) time scale. In our case, the quartz clock of each recorder module plays the role of a local scale, which is periodically checked against the UTC scale according to the algorithm described in chapters \ref{Observations},\ref{Clocks}.
\item Takes into account the corrections of the local time scale relative to the external scale (UTC) and recalculates the TOA to the UTC time scale.
\item Reduction of topocentric TOA of the pulsar under study into the barycentric coordinate system of the Solar system and barycentric time (obtaining barycentric TOA). Obtaining barycentric residual deviations (TOA) as the differences (inconsistencies) between the observed and calculated barycentric TOA based on pulsar ephemerides.
\item Refinement of pulsar parameters in a barycentric reference frame by the method of successive approximations according to the criterion of maximum likelihood (minimizing the frequency response of barycentric TOA over the observation interval).
\end{itemize}

It is easy to see that in our case, the main problem in implementing the algorithm is the second point related to the time binding of the local standard to the UTC time scale. As practice has shown (see Chapter \ref{Clocks}), quartz clocks can go off during the time between synchronization with the external scale, reaching values much higher than acceptable for timing, in particular, exceeding the period of the observed pulsar (i.e., there is a loss of pulse phase). At the same time, hardware binding, especially in the early stages of observations, up to approximately MJD = 57500, could be carried out irregularly and the total departure of the local standard relative to UTC could be $\gg 12.4928$~ms, while the departure of the local standard one hour after synchronization, as noted above, is the value of $< 2$ points.

To calculate the corrections of the quartz clocks of each module relative to the UTC scale for each day of observations, we used an independent intermediate astronomical pulsar time (PT), implemented by an ensemble of powerful second pulsars with well-known parameters, observed daily on LPA3 in the background. The main stages of the practical implementation of this algorithm will be described in detail below.

\subsection{Accumulation of the mean profile of the pulsar} \label{Integration}

Due to the movement of the reference clock, the observed phase of the pulsar pulse within the time window equal to its current period will shift relative to the calculated one. A shift in the observed phase can have consequences when averaging the pulsar signal over long time periods, since when adding pulses, for example, over two or more consecutive days, the signal will not add up in phase. This will lead to an expansion/disappearance of the mean profile, and it will not be possible to achieve the expected increase in the S/N ratio in the mean profile. However, if the clock error during one observation session, which is 3.5 minutes, is less than 1 point (frequency oscillator count = 12.4928 ms), then the signal averaging for it is coherent - the pulsar signal is isolated from the noise and can be reliably measured.

It should be noted that for timing, the criterion for maintaining the observed phase of a pulsar with accuracy to a certain degree is quite strict. If the S/N ratio is satisfactory for slow pulsars (the profile width of which is $\ll P$) and if the purpose of timing is to calculate the pulsar ephemeris satisfactory for further observations, a soft criterion can be used when phase conservation is allowed within $\le W_{0.5}$ (i.e., a possible expansion is allowed the average profile is up to its width by 1/2 of the peak flux density).

We tested the possibility of accumulating an mean profile of satisfactory quality for timing over a time period of about 3 years. At the same time, the "blurring" of the pulse when it is summed up in this time interval is influenced by both the course of the local clock and the inaccuracy of the catalog parameters. In the example discussed below, the pulsar parameters are known with high accuracy, and the movement of quartz clocks had a dominant effect on the distortion of the pulse shape.

\begin{figure}
\includegraphics[width=\columnwidth]{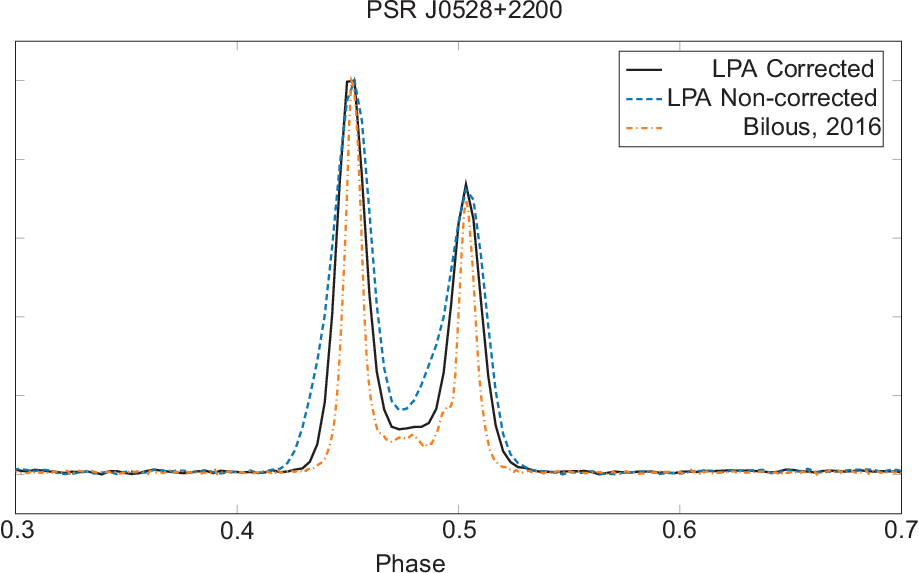}
\caption{The mean profile of the pulsar J0528+2200, obtained by adding 1000 sessions taken over an interval of 3 years. On the abscissa axis is the pulse phase, on the ordinate axis is the normalized flux density (in 
conventional units). The black solid line shows the profile obtained at the LPA after adjusting the clock, compiled
taking into account the parameters from the ATNF catalog. The blue dotted line shows the same profile before 
 adjusting the clock. For comparison, the orange dotted line shows the profile obtained in 
(\citeauthor{Bilous2016}, \citeyear{Bilous2016}).}
\label{Fig1}
\end{figure}

Fig.\ref{Fig1} shows an example of an mean profile obtained by adding the pulses of a strong pulsar J0528+2200 (B0525+21) with a two-component profile over 1000 observation sessions. When adding up, the expected pulse phase (in an ideal model) was calculated for each day of observations using the TEMPO2 phase analysis program (\citeauthor{Hobbs2006}, \citeyear{Hobbs2006}) based on the model parameters: $P, \dot P$ - the period of the pulsar's own rotation and its first time derivative (or alternatively $\nu$ and $\dot \nu$ - frequency and its derivative), $\alpha, \delta, \mu\alpha, \mu\delta$ (coordinates and proper motion), $DM$ (dispersion measure) from the ATNF catalog. In the case of an ideal clock and accurate pulsar parameters, an average profile should be obtained, the distortion (broadening) of the shape of which does not exceed the sampling (12.4928 ms). The width of the pulsar profile is $W_{0.5} \approx$230~ms at frequencies of 102.75 and 119.72 MHz (\citeauthor{Kuzmin1999}, \citeyear{Kuzmin1999}; \citeauthor{Smirnova2009}, \citeyear{Smirnova2009}; \citeauthor{Bilous2016}, \citeyear{Bilous2016})\footnote{https://psrweb.jb.man.ac.uk/epndb /}, that is, at the frequency of LPA observations equal to 
110.25 MHz, we can expect a close value of $W_{0.5}$.

The graph in Fig.\ref{Fig1} shows that the mean profile obtained by us over the interval of 3 years (1000 days) has $W_{0.5} = 261$~ms, and the profile structure remains two-component. This example shows that monitoring data obtained on the LPA3 can be used to obtain mean pulsar profiles if the pulsars have well-defined values of $P, \dot P$ and coordinates. The additional pulse broadening was 261-230 = 31 ms $\approx 2.5$ points. This difference in the half-widths of the mean profile demonstrates the imperfection of the local clocks used in his observations. In the same figure (somewhat looking ahead - see sections 4.2 and 4.3 of Chapter IV), we also presented the mean profile after taking into account the clock movement of quartz oscillators according to the developed algorithm. It can be seen from the figure that the mean profile obtained on LOFAR is the narrowest, and the profile width occupies an intermediate position after correction of the clock ($W_{0.5}$=244 ms). Thus, after taking into account the clock's progress, the difference between our and the best (LOFAR) pulsar profile was only $244 - 231=13$~ms$\approx 1$ point, which is satisfactory and meets our expectations.

\subsection{Refinement of pulsar parameters} \label{Timing.Fitting}

As noted above, during the timing process, topocentric TOAs are reduced to barycentric ones, for which the astrometric and rotational parameters of the pulsar, the parameters of the interstellar medium, and in the case of a binary pulsar, its orbital parameters are used. In our task, the number of parameters is limited due to the fact that the vast majority of slow pulsars are single (the ATNF catalog contains 32 double pulsars with $P>100$ ms). The standard model of a single pulsar includes the period and its time derivatives $P, \dot P, \ddot P$, coordinates and proper motions of the pulsar $\alpha,\delta, \mu\alpha, \mu\delta$, parallax $\pi$, $DM$. This set of parameters makes it possible to determine the propagation delays of a signal in the Solar System in the direction of the observer at any time of observation. Geometric and relativistic corrections related to the propagation of the pulsar pulse in the gravitational field of the Sun, Moon, and large planets of the Solar System are parameterized by the relative position of the pulsar and Solar System bodies on the celestial sphere calculated from the model at a given moment in time, calculated from ephemerides (for example, DE405 or DE440).
Limitations on the accuracy of determining the TOA, determined by the movement of quartz clocks in the hourly interval of observations, force us to limit ourselves to the parameters $P, \dot P$, sometimes $\ddot P$ (or, equivalently, $\nu, \dot\nu$) $\alpha, \delta$ is visible. Since the observations are carried out at a single frequency in a narrow band ($\approx 2.5$~MHz), the parameter $DM$ cannot be independently determined in our observations and must be set externally. Throughout, we used the values of $DM$ from the ATNF pulsar catalog, considering this value to be a constant for the entire time of observations.

The topocentric TOA of a pulsar in a session was determined by us as the sum of the moment of the beginning of the observation session and the moment of time tied to the reference point on the mean profile obtained by averaging individual pulses during one observation session. The reference point is defined as the maximum of the cross-correlation function with the template pulse profile of the same pulsar (\citeauthor{Doroshenko1990}, \citeyear{Doroshenko1990}). The template pulse profile, in turn, is obtained by summing the mean profiles obtained in a series of observations, while preserving the phase. The calculation of template profiles and topocentric TOAs was performed using the AnTi-pipeline program\footnote{https://github.com/StepanRepo/AnTi-pipeline} and Python3 libraries: skyfield (\citeauthor{Rhodes2019}, \citeyear{Rhodes2019}), astropy (\citeauthor{Price-Whelan2022}, \citeyear{Price-Whelan2022}), numpy (\citeauthor{Harris2020}, \citeyear{Harris2020}), scipy (\citeauthor{Virtanen2020}, \citeyear{Virtanen2020}).

The TOA was reduced to the barycenter of the Solar System and the barycentric time, as well as the parameters of the pulsar were refined using the TEMPO2 program. Based on the specified initial parameters of the pulsar from the catalog, model barycentric TOA and residuals - modeling inconsistencies equal to the difference between the observed and calculated TOA were calculated. The residuals structure allows us to judge the type of modeling errors: an error in the value of one of the parameters or a physical effect that was not taken into account in the model. In particular, the residuals obviously contains unaccounted for corrections of the local clock. The residuals of the model, which fully describes the observed TOA series, has the form of white phase noise.

\subsection{Construction of corrections to the local scale and the structure of the residual deviations of the TOA} \label{Timing.Correction}

The general algorithm for constructing corrections in the residuals TOA and taking them into account when calculating the refined TOA and pulsar parameters is as follows:

\begin{itemize}
\item Calculation of barycentric TOA (reduction) and residuals TOA for all reference pulsars and each module with reliable initial parameters. Subtraction of the harmonic component from the residuals of each pulsar due to an inaccurate initial value of its coordinates and/or proper motions.
\item Calculation of the time correction for each of the modules. As such, the general component of the residuals is assumed, mainly due to the course of the local clock. The correction for each date is calculated as the median value of all pulsars of the module on that day.
\item Calculation of the barycentric TOA of the pulsar under study observed in this module and their correction over the course of the local clock: subtracting the corrections calculated at the previous stage from them.
\item Refinement of the parameters of the pulsar under study using the TOA corrected over the course of the clock.
\end{itemize}

\begin{figure*}
\includegraphics[width=0.7\textwidth]{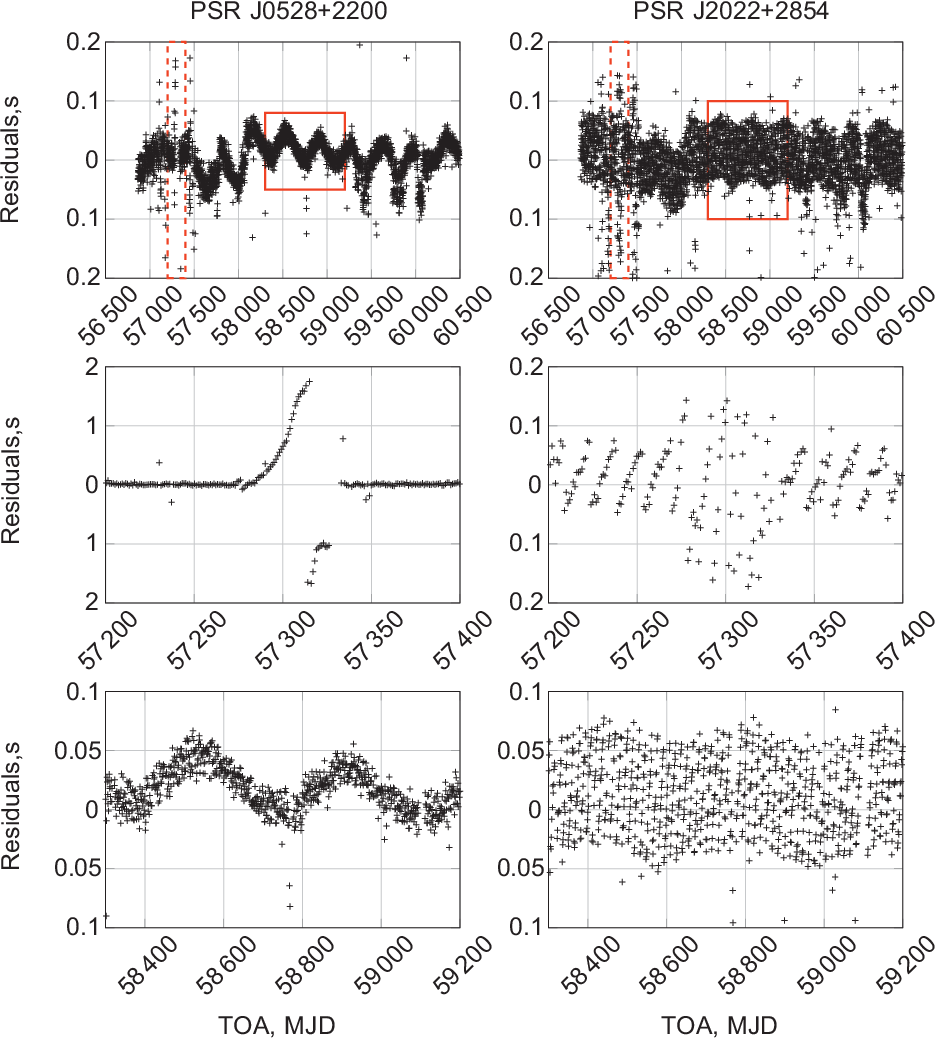}
\caption{Residual deviations of the TOA of pulsars J0528+2200 and J2022+2854. The TOA is shown along the abscissa axis, and the TOA is shown along the ordinate axis (modulo the pulsar period within $\pm 1/2 P$). The upper figures show the general view of the residuals over the entire observation segment. Dotted lines indicate the areas shown separately in the second row, which show outliers associated with a synchronization error of the local standard. Solid rectangles highlight the sections shown separately in the bottom line, on which the harmonic component of the residual is clearly visible.}
\label{Fig2}
\end{figure*}

Note that during the operation of the algorithm, the rotational parameters of the reference pulsars are not specified in the first step according to the r.m.s., since the linear and quadratic trends of the reference clock would inevitably be distorted if the parameters of the pulsar period and its derivatives were changed. At the same time, the harmonic component of the residual can be identified and taken into account quite confidently, at least in part of the data (see graphs in Fig.\ref{Fig2}).

We further believe that the residuals caused by the inaccuracy of the values of the period and its derivatives for different pulsars of each module do not correlate with each other and are sufficiently well eliminated by averaging the TOA series. This is equivalent to assuming that, within the range of errors caused by "white noise", or residual noise, the corrections are due only to the movement of the reference (quartz) clock. Such an assumption is quite possible, given that the purpose of our work is, first of all, to obtain not the best possible, but a sufficiently reliable initial set of parameters for new pulsars, which will allow them to continue precision timing at high radio frequencies.

We selected 24 reference pulsars to make corrections to the local time scale and verify the possibility of timing with poor local time reference. The selection criteria were: the value of the period in the catalog, known with an accuracy of at least $10^{-10}$~s; the pulsar contains in the studied area of the sky ($+21^{\circ} < \delta < +42^{\circ}$); good visibility of the pulsar in a separate observation session (S/N mean profile $\ge 7$). At this stage, the values of the pulsar parameters given in the ATNF were taken as the initial parameters for TOA reduction, which were considered to be {\it a priori} accurate.
Fig.\ref{Fig2} shows the residuals we obtained for two pulsars (before applying the algorithm), which were detected by different (fourth and sixth, respectively) modules of the LPA3 radio telescope recorder. A check on all pulsars showed that the residuals of objects observed in the same modules exhibit the same structures, and the structure of the residuals of pulsars in different modules has, along with this, its own characteristics. This is obviously due to the fact that each module, as mentioned above, has its own reference quartz oscillator with individual frequency departure and relative error, as well as a general reference to an external time scale.

The residual structure shown in Fig.\ref{Fig2} can be divided into several components: outliers, large-scale, "irrational" a component, a harmonic component with a period of 1 year, and a series of periodic straight segments with a period of 15 days. Moreover, outliers, irrational and harmonic components are repeated for all studied objects, and periodic straight line segments have the same character for objects of each module.
Let's explain the nature of the observed effects in more detail:

\begin{enumerate}

\item Large-scale emissions caused by violations of the synchronization procedure of the local standard (quartz watch) with the UTC scale. In the intervals where outliers are visible, the local standard has not been compared for a long time. Since the frequency of the local standard stated and indicated in the observational data file differs from its actual frequency, the time scale associated with it accumulated a difference with the UTC scale. During the period MJD $\approx 57275-57325$ in Fig.\ref{Fig2} residual noises are visible (limited by dashed red lines), which are noticeably higher than in other observed periods. This is because during this period, due to the large synchronization interval, the discrepancy between the registrar's system time scale and the NTP server scale (linked to the UTC scale) could reach units of seconds for some modules. Subsequently, the synchronization interval of the system time of the recorders and the time scale of the NTP server was reduced to 1 hour. During the operation of the recorders, there were also cases of unstable operation of the local network, which could also lead to a loss of synchronization (MJD $\approx$59400, 59800, 60100). In the middle row in Fig.\ref{Fig2} shows clearly visible on the graph for J0528+2200, the maximum of the observed emissions, which reached a difference from UTC of almost 2 seconds. For the pulsar J2022+2854, the so-called "phase loss" is observed in this time interval, i.e. residuals more than $P$.

\item The harmonic component is shown in the lower panel of Fig.\ref{Fig2}. This component is most strongly manifested in three objects: J0528+2200, J0629+2415, J0826+2637. Such a structure in the pulsar's residuals usually indicates insufficient accuracy of the initial coordinates given in the ATNF catalog, which should manifest itself in the appearance of a sinusoid, the phase of which is associated with the ecliptic longitude, and the amplitude with the ecliptic latitude of the pulsar. This effect is associated with the peculiarities of determining the coordinates of pulsars by the method of timing, and usually manifests itself in pulsars with a strong polynomial course of the TOA of an unknown nature (the so-called "pulsar noise"), whose coordinates were determined at time intervals of less than one and a half years. In this case, the corrections in the TOA related to the coordinates of the pulsars and the derived periods of the pulsar become indistinguishable, and when the duration of observations changes, the estimates of these parameters can be redefined. We also note that due to the mutual influence of these factors and the presence of polynomial moves of the reference standard, the annual sinusoid cannot always be completely eliminated from the data, which, in turn, may lead to a slight (within the limits of satisfactory accuracy) redefinition of the coordinates of the pulsars observed within this module.

\item Periodic straight lines, clearly visible on the right middle panel of Fig.\ref{Fig2}, they appear the same for objects inside the same module. This error is the effect of comparing two frequency generators: the frequency grid of the module and the UTC scale. When the two scales are synchronized, corrections are made to the local scale, which make up the whole number of oscillation of the generator. As time passes, the difference between the scales accumulates until it reaches one whole oscillation, which is subtracted in the next synchronization. This synchronization helps to keep the running error of the local standard limited.
\end{enumerate}

In the future, when timing weak pulsars and RRATs, whose average profile or pulses are only rarely visible, we can rely on uniform corrections in each module of the recorders. Thus, corrections for the course of the local scale based on observations of reference pulsars will make it possible to use quartz clocks as reference local clocks that are stable over long periods of time, with the maximum limitation on the accuracy of determining the TOA for which the best time resolution of observations will be.

\subsection{Estimation of the accuracy of parameter detection} \label{Timing.Accuracy}

\begin{figure*} 
\includegraphics[width=0.7\textwidth]{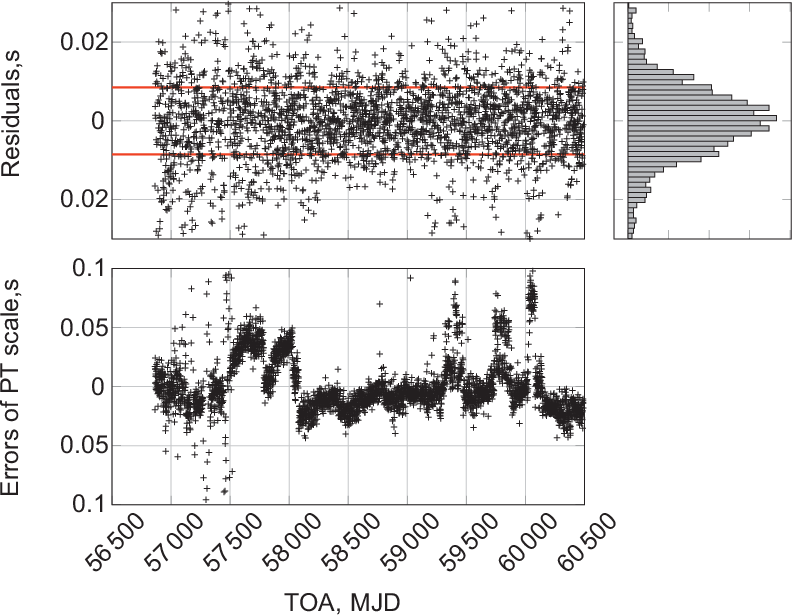} 
\caption{On the upper panel: the TOA of the pulsar J1921+2153, obtained using the parameters from the Table~1 and Table~2 after taking into account the corrections of the reference clock. The vertical axis shows the values of residuals in seconds, and the horizontal axis shows the TOA in MJD. The horizontal lines show the level of the r.m.s.. The right panel shows the distribution of the residuals TOA over the pulse phase. It can be seen that the residuals distribution corresponds well to the model of white (Gaussian) phase noise. On the bottom panel is the correction calculated by us, compensating for the movement of the clock of the 6th module, used in the correction of the TOA.} 
\label{Fig3} 
\end{figure*}

As a result of taking into account the effects of the course of the local frequency standard and clarifying the rotational and astrometric parameters, we were able to bring the residuals of the pulsars of our sample to a noise track having a form close to white phase noise. In Fig.\ref{Fig3} shows an example of an residuals for pulsar J1921+2153 (B1919+21) after using corrections to adjust the module's local clock. The pulsar is located in the 6th module of the recorder, which contains 7 reference pulsars. The magnitude of the errors is consistent with their estimate given in (\citeauthor{Lorimer2008}, \citeyear{Lorimer2008}). We note the quasi-periodic variations with a period of about a year, visible on the graph of the residual TOA corrections (Fig.\ref{Fig3}, bottom panel). The reason for their occurrence is given in the previous section of this chapter.

For the entire sample of pulsar frequency response data (taken at intervals of about a month and not including large outliers) before calibration of the local standard, it ranged from 20 to 100 ms, depending on the quartz oscillator (frequency grid) and the corresponding module. After making corrections, the operating time of 24 reference pulsars does not exceed 20 ms for all modules. For comparison, typical residuals in conventional timing are usually units of milliseconds for slow pulsars (after reducing residuals to "white noise"). We have selected a network of reference pulsars so that each module has at least one such pulsar. In the future, the number of reference pulsars in the module is expected to be increased to eliminate self-calibration and introduce systematic errors in the determination of the period and its derivatives in new pulsars. Usually, 3-4 pulsars in different beams of LPA3 are observed in the module. 

%\begin{figure*} 
%\includegraphics[width=0.7\textwidth]{fig3eng.eps} 
%\caption{On the upper panel: the TOA of the pulsar J1921+2153, obtained using the parameters from the Table~1 and Table~2 after taking into account the corrections of the reference clock. The vertical axis shows the values of residuals in seconds, and the horizontal axis shows the TOA in MJD. The horizontal lines show the level of the r.m.s.. The right panel shows the distribution of the residuals TOA over the pulse phase. It can be seen that the residuals distribution corresponds well to the model of white (Gaussian) phase noise. On the bottom panel is the correction calculated by us, compensating for the movement of the clock of the 6th module, used in the correction of the TOA.} 
%\label{Fig3} 
%\end{figure*} 

The results of applying the method to a sample of all 24 pulsars can be seen in Tables \ref{tab1} and \ref{tab2}, which show their r.m.s. refined rotational parameters and coordinates in comparison with data from the ATNF catalog. 

The Table\ref{tab1} shows the results of the evaluation of the rotational characteristics of the reference objects according to the monitoring data of the LPA. The objects are arranged in descending order of declination. Column 2 of the table shows the name of the pulsar. Columns 3-6 show: the frequency of the pulsar's own rotation $\nu_{obs}$, estimated using the TEMPO2 program according to monitoring data after taking into account the movement of the quartz clock, the r.m.s. estimate of its error is indicated in parentheses.; the difference between the pulsar rotation frequencies according to our data and the ATNF catalog: $\nu_{obs}-\nu_{ATNF}$, in cases where the difference is less than the error in determining the corresponding parameter, a dash is indicated in the table; the derivative of the rotation frequency obtained according to our data is $\dot\nu_{obs}$, the difference between the frequency derivatives in this work and in the ATNF catalog $\dot\nu_{obs}- \dot\nu_{ATNF}$. The last column shows the standard deviation level (r.m.s.) of the TOA for all objects. The values in columns 4-6 are accurate to the last significant digit. 

\begin{table*} 
\caption{Estimates of the rotational parameters of 24 reference pulsars}
	\begin{tabular}{|r|r|r|r|r|r|r|}
		\hline
		\rot{Modulus} & 
		\rot{JNAME} & 
		\rot{$\nu_{\text{obs}}$ s$^{-1}$} & 
		\rot{$\nu_{\text{obs}} - \nu_{\text{ATNF}}$, $10^{-10}$ s$^{-1}$ } & 
		\rot{$\dot {\nu}_{\text{obs}}$, $10^{-15}$ s$^{-2}$} & 
		\rot{$\dot{\nu}_{\text{obs}} - \dot{\nu}_{\text{ATNF}}$, $10^{-18}$ s$^{-2}$} &
		\rot{r.m.s. residuals,\ $10^{-3}$ s}\\
		%\midrule
		\hline
		1 & J1821+4147  & $0.792\,482\,692\,772(27)$ & $8$    & $-1.081$   & $2  $  & 11 \\
		&J2208+4056  & $1.569\,963\,720\,061(54)$ & $13  $ & $-13.02$ & $7  $  & 12 \\
		&J1813+4013  & $1.074\,011\,087\,813(39)$ & $-10 $ & $-2.945$ & $-   $  & 10 \\
		&J1907+4002  & $0.809\,220\,283\,982(59)$ & $25  $ & $-0.351$ & $-1  $  & 14 \\
		&J2157+4017  & $0.655\,623\,511\,13(53)$  & $-68 $ & $-1.502$ & $-29$  & 14 \\
		&J0323+3944  & $0.329\,807\,475\,488(15)$ & $9   $ & $-0.068$ & $-   $  & 11 \\
		\hline
		2 & J0613+3731  & $1.614\,991\,846\,387(13)$ & $-222$ & $-8.439$ & $-    $  & 7  \\
		& J0612+3721  & $3.355\,903\,712\,29(11)$ & $136 $ & $-0.648$ & $-    $  & 9  \\
		& J0612+37216 & $2.252\,904\,945\,211(15)$& $17  $ & $-0.779$ & $-    $  & 7  \\
		\hline
		3 & J0048+3412  & $0.821\,629\,023\,056(65)$ & $21  $ & $-1.586$ & $3  $  & 22* \\
		\hline
		4 & J2305+3100  & $0.634\,563\,531\,351(25)$ & $1$ & $-1.166$ & $1    $  & 8  \\
		& J2018+2839  & $1.792\,264\,110\,834(61)$ & $27 $ & $-0.475$ & $-   $  & 6  \\
		& J2022+2854  & $2.912\,037\,609\,2(13)$  &  $42 $ & $-16.07$ & $-30$  & 8  \\
		\hline
		5 & J1532+2745  & $0.889\,018\,677\,71(70)$ & $136 $ & $-0.586$ & $26 $  & 15 \\
		& J2113+2754  & $0.831\,357\,631\,0(10)$ &  $127 $ & $-1.786$ & $24 $  & 19 \\
		& J0826+2637  & $1.884\,444\,020\,6(28)$ &  $-532$ & $-6.212$ & $-262$ & 15 \\
		& J1239+2453  & $0.723\,353\,914\,0(12)$ &  $220 $ & $-0.463$ & $36 $  & 18 \\
		\hline
		6 & J0629+2415  & $2.098\,095\,022\,550(92)$ & $2   $ & $-8.779$ & $-1   $  & 12 \\
		& J0943+2253  & $1.876\,261\,645\,492(57)$ & $30  $ & $-0.310$ & $-1  $  & 13 \\
		& J2055+2209  & $1.226\,721\,269\,532(48)$ & $33  $ & $-2.012$ & $-   $  & 11 \\
		& J0528+2200  & $0.266\,984\,252\,752(7)$ &  $6   $ & $-2.853$ & $1    $  & 13 \\
		& J1238+2152  & $0.893\,982\,048\,236(29)$ & $39  $ & $-1.146$ & $-1   $  & 11 \\
		& J1921+2153  & $0.747\,774\,158\,234(27)$ & $21  $ & $-0.752$ & $-    $  & 9  \\
		& J2317+2149  & $0.692\,207\,697\,099(31)$ & $20  $ & $-0.499$ & $-   $  & 11 \\
		\hline
		%\bottomrule
	\end{tabular}
	\label{tab1}		
\end{table*}
Note: The J0048+3412 object is the only one in the module, as a result, the TOA was obtained after making corrections to the TOA only for the most obvious polynomial moves of the clock, this was reflected in their relatively large value for the r.m.s. residuals TOA.

It can be seen from the Table\ref{tab1} that estimates of the pulsar rotation frequency differ from the values given in the ATNF, typically by $10^{-8} - 10^{-9}$~Hz. Assuming that the catalog parameters are accurate, translated into periods, this will mean that the "shift" period occurs approximately after $10^8-10^9$ revolutions of the pulsar. For example, for a pulsar with a period of $P_0=1$~s "shift" of the period (accompanied by phase loss) will occur in the range from 3 to 10 years.

The Table\ref{tab2} shows the results of estimating the coordinates of the reference objects according to the LPA monitoring data, Column 2 of the table shows the name of the pulsar. Columns 3-4 show the estimates of coordinates obtained in the course of the work for right ascension and declination, $\alpha_{obs}$ and $\delta_{obs}$, respectively. Columns 5-6 show estimates of coordinate errors in the ATNF catalog, $\sigma\alpha_{ATNF}$ and $\sigma\delta_{ATNF}$. Columns 7-8 show parameter errors estimated as r.m.s. $\sigma\alpha_{obs}$ and $\sigma\delta_{obs}$. Columns 9-10 show the difference between the estimates of right ascension and declination obtained in the course of this work and those given in the ATNF catalog, $\alpha_{obs}-\alpha_{ATNF}$ and $\delta_{obs}-\delta_{ATNF}$.

\begin{table*}
\caption{Estimates of the coordinates of 24 reference objects on the celestial sphere}
	\begin{tabular}{|r|r|r|r|r|r|r|r|r|r|}
		\hline
		\rot{Modulus} &
		\rot{JNAME} & 
		\rot{$\alpha_{\text{obs}}$, hh:mm:ss  } & 
		\rot{$\delta_{\text{obs}}$, ${}^{\circ}:':''$} & 
		\rot{$\sigma \alpha_{\text{ATNF}}$, $10^{-3}$s} & 
		\rot{$\sigma \delta_{\text{ATNF}}$, $10^{-3}$ $''$} & 
		\rot{$\sigma \alpha_{\text{obs}}$, $10^{-3}$ s} & 
		\rot{$\sigma \delta_{\text{obs}}$, $10^{-3}$ $''$} & 
		\rot{$\alpha_{\text{obs}} - \alpha_{\text{ATNF}}$, s} &
		\rot{$\delta_{\text{obs}} - \delta_{\text{ATNF}}$, $''$}\\
				%\midrule
		\hline
		1 & J1821+4147  & 18:21:52.25 & +41:47:03.2 & 4.1  & 36   & 10 & 118  & $-0.05$ & $0.7 $  \\
		& J2208+4056  & 22:08:01.84 & +40:55:59.8 & 9.8  & 130  & 11 & 139  & $-0.06$ & $-2.0$  \\
		& J1813+4013  & 18:13:13.12 & +40:13:40.2 & 7.0  & 110  & 9  & 114  & $-0.08$ & $1.2 $  \\
		& J1907+4002  & 19:07:34.56 & +40:02:06.4 & 8.0  & 110  & 16 & 207  & $-0.04$ & $0.7 $  \\
		& J2157+4017  & 21:57:01.66 & +40:17:44.2 & 0.1  & 2    & 14 & 177  & $-0.19$ & $-1.7$  \\
		& J0323+3944  & 03:23:26.28 & +39:44:50.3 & 0.1  & 1    & 12 & 321  & $-0.38$ & $-2.1$  \\
		\hline
		2 & J0613+3731  & 06:13:11.89 & +37:31:42.7 & 11.0 & 900  & 6  & 304  & $-0.25$ & $4.4 $  \\
		&J0612+3721  & 06:12:48.44 & +37:21:42.0 & 1.5  & 110  & 8  & 369  & $-0.25$ & $4.7 $  \\
		&J0612+37216 & 06:12:43.86 & +37:21:46.8 & 30.0 & 2200 & 6  & 312  & $-0.24$ & $6.8 $  \\
		\hline
		3 & J0048+3412  & 00:48:33.44 & +34:12:04.5 & 21.0 & 370  & 21 & 422  & $-0.54$ & $-3.5$  \\
		\hline
		4 & J2305+3100  & 23:05:58.24 & +30:59:58.0 & 0.1  & 1    & 9  & 164  & $-0.08$ & $-3.3$  \\
		& J2018+2839  & 20:18:03.75 & +28:39:54.6 & 0.9  & 16   & 6  & 91   & $-0.08$ & $0.4 $  \\
		& J2022+2854  & 20:22:37.09 & +28:54:23.5 & 0.7  & 11   & 7  & 119  & $-0.08$ & $0.4 $  \\
		\hline
		5 & J1532+2745  & 15:32:10.27 & +27:45:50.8 & 0.1  & 1    & 12 & 206  & $-0.10$ & $1.2 $  \\
		& J2113+2754  & 21:13:04.15 & +27:53:57.0 & 0.1  & 1    & 16 & 270  & $-0.15$ & $-4.1$  \\
		& J0826+2637  & 08:26:51.33 & +26:37:34.1 & 0.1  & 1    & 24 & 1182 & $-0.17$ & $12.8$  \\
		& J1239+2453  & 12:39:40.33 & +24:53:54.1 & 1.2  & 20   & 18 & 386  & $-0.14$ & $4.8 $  \\
		\hline
		6 & J0629+2415  & 06:29:05.42 & +24:16:00.2 & 0.1  & 1    & 26 & 6759 & $-0.31$ & $18.6$  \\
		& J0943+2253  & 09:43:32.14 & +22:53:10.2 & 1.0  & 40   & 25 & 898  & $0.06 $ & $6.8 $  \\
		& J2055+2209  & 20:55:38.93 & +22:09:24.5 & 4.5  & 100  & 9  & 171  & $-0.22$ & $-2.7$  \\
		& J0528+2200  & 05:28:51.99 & +21:59:27.5 & 9.0  & 2000 & 27 & 6525 & $-0.27$ & $-36.5$ \\
		& J1238+2152  & 12:38:22.87 & +21:52:14.2 & 78.0 & 1400 & 11 & 252  & $-0.30$ & $3.1 $  \\
		& J1921+2153  & 19:21:44.59 & +21:53:00.5 & 2.0  & 40   & 7  & 134  & $-0.22$ & $-1.7$  \\
		& J2317+2149  & 23:17:57.53 & +21:49:44.8 & 0.1  & 1    & 11 & 253  & $-0.32$ & $-3.2$  \\
		%\bottomrule
		\hline
		
	\end{tabular}
	\label{tab2}		
\end{table*}

As can be seen from the Tables, the typical estimates of accuracy (errors) obtained by our method after clarifying the parameters according to the r.m.s. are in the ranges: $\delta \nu_{err} \sim 10^{-9} - 10^{-11}$~ s$^{-1}$; $\delta \dot \nu_{err} \sim 10^{-14} - 10^{-16}$~ s$^{-2}$; $\alpha_{err} \sim 0.1^s$; $\delta_{err} \sim 0.1 -10^{\prime \prime}$.

The largest error in determining the period may differ by 1-2 orders of magnitude from the difference between the periods defined by us and in the ATNF catalog. This may be due to several factors, (1) the model we used does not fully compensate for the polynomial moves of the TOA (time polynomials of degree higher than $t^2$) associated with the movement of quartz standards. These moves lead to redefinition of the proper rotation parameters. (2) For a number of pulsars, the parameters in the ATNF catalog were determined at shorter time intervals than in our paper (for example, surveys of LOFAR or MeerKAT). The determination of the parameters given in the catalog on shorter time intervals than ours also led, apparently, to significant differences in the estimates of declination and right ascension (see for more information the reasons for the occurrence of the annual sinusoid in the TOA and the offset of the estimates of coordinates in the Section\ref{Timing.Correction}).

\section{Discussion of the results and conclusion} \label{Discussion}

We have developed and tested a method for timing slow pulsars using monitoring data from the northern sky in the declinations $+21^{\circ} < \delta < +42^{\circ}$, held for 10 years. The method makes it possible to circumvent the limitations on determining the TOA of pulsars on a local scale, imposed by the low accuracy of the quartz watches used to form it. For this purpose, groups of stable slow pulsars with well-known parameters are used as an intermediate reference time standard. The accuracy of determining the rotational parameters and coordinates of pulsars achieved in this way on a sample of 24 pulsars is about two orders of magnitude worse than the accuracy of timing slow pulsars achieved to date in the meter wavelength range (\citeauthor{Tan2020}, \citeyear{Tan2020}; \citeauthor{Potapov2023}, \citeyear{Potapov2023}). Nevertheless, the calculated set of parameters can be used in phase analysis programs (TEMPO2, TIMAPR405) as a starting point for making observations according to the high-frequency timing program. The achieved accuracy is sufficient for use in a number of tasks where high accuracy in determining TOA is not required, but it is necessary to ensure the possibility of accumulating the pulsar pulse during the observation session (see Introduction).

For example, such a task, where high accuracy is not required, may be the task of confirming Pushchino new pulsars and RRATs by detection on other radio telescopes. The typical dimensions of the radiation pattern in observations at 1.4 GHz on large radio telescopes are $5-10^\prime$, the angular size of the FAST beam reaches $3^\prime$. When independently checking pulsars, problems arise related to the loss of time both for searching different directions in the sky for the initial detection of the pulsar, and with the accumulation of a pulse with a poorly known period (which is especially critical when using the synchronous pulse accumulation method for observations). Considering that the typical coordinate error on the LPA is $\pm 15'$, it is necessary to spend 5-10 times more observational time than for the case of coordinates known with minute and sub-minute accuracy. This work shows that using the coordinates obtained by the proposed method, it will be possible to immediately include new pulsars and RRATs discovered at the LPA in research programs on large telescopes with a narrow directional pattern.

Note that currently in the list of pulsars discovered on the LPA\footnote{https://bsa-analytics.prao.ru/pulsars/new /} there are 87 objects whose periods are accurately determined. $10^{-3} - 10^{-4}$~s, which are candidates for parameter refinement by this method. We also note that the ATNF catalog included 3678 pulsars at the beginning of 2025. For about 600 pulsars, the period is known with an accuracy of no better than the 6th, and for $\sim$300 - no better than the 4th decimal place. Some of these pulsars are detected at LPA3 and their parameters can be refined.

The PUMPS project has been under observation at the LPA for more than 10 years. Almost 100 RRATs were detected in these observations, and the period of these transients is known with no better accuracy than 3-4 decimal places, and often not at all. Confirming the detected RRATs on other radio telescopes, taking into account the appearance of pulses sometimes every few hours and a large coordinate error, becomes a very difficult task. That is, the study of the RRAT detected in Pushchino, given the accuracy of the coordinates that we give, is almost impossible on other large telescopes. However, if the coordinate is obtained with an accuracy of $10^{\prime\prime}$, with a known $DM$, other telescopes can be used to study them. In addition, timing will make it possible to determine the position of the RRATs open in the PRAO on the $P/\dot P$ diagram and determine the physical characteristics of the observed transients based on the mechanism of magnetodipole radiation. In other words, timing will allow us to test hypotheses about the nature of the RRAT.

\section*{ Acknowledgements}
We would like to thank the antenna group serving the LPA for constantly monitoring its operability and timely repair. S.A.T. expresses its gratitude to the grant of the Russian Science Foundation 25-12-68023 (https://rscf.ru/project/25-12-68023 /) for supporting the work on transient research.

%\bibitem{Coenen2014} Coenen, T., van Leeuwen, J., Hessels, J.~W.~T., et al., A\&A, {\bf 570}, A60, (2014). 

\end{document}